# Increased formation of trions and charged biexcitons by above-gap excitation in single-layer $WSe_2$

*Matthew C. Strasbourg[1,2]†, Emanuil S. Yanev[2], Sheikh Parvez[1], Sajia Afrin[3], Cory Johns[1], Zoe Noble[1], Thomas P. Darlington[2], Erik M. Grumstrup[3], James C. Hone[2], P. James Schuck[2], and Nicholas J. Borys[1]**

[1]Department of Physics, Montana State University, Bozeman, MT

[2]Department of Mechanical Engineering, Columbia University, New York, NY

[3]Department of Chemistry, Montana State University, Bozeman, MT

**Abstract**

Two-dimensional semiconductors exhibit pronounced many-body effects and intense optical responses due to strong coulombic interactions. Consequently, subtle differences in photoexcitation conditions can strongly influence how the material dissipates energy during thermalization. Here, using multiple excitation spectroscopies, we show that a distinct thermalization pathway emerges at elevated excitation energies, enhancing the formation of trions and charged biexcitons in single-layer $WSe_2$ by up to 2× and 5×, respectively. Power- and temperature-dependent measurements lend insight into the origin of the enhancement. These observations underscore the complexity of excited state relaxation in monolayer semiconductors, provide insight for the continued development of carrier thermalization models, and highlight the potential to precisely control excitonic yields and probe non-equilibrium dynamics in 2D semiconductors.

Two-dimensional (2D) single-layer (1L) transition metal dichalcogenide (TMD) semiconductors are rich excitonic systems for investigating fundamental light-matter interactions[1-5], harvesting light on the nanoscale[6, 7], generating quantum states of light[8-10], biosensing[11-14], and probing non-equilibrium physics[15, 16]. In 1L-$WSe_2$, a prototypical 2D semiconductor, a manifold of optically emissive excitonic states underpins these capabilities and dominates photoluminescence processes. The manifold consists of bright and dark[17-19] excitons, biexcitons[20-22], trions[23], and charged biexcitons[24-27]. In addition, higher energy excitons and free-carrier states exist[28-30]. Whereas many of these additional states efficiently absorb light, they exhibit substantially weaker photoluminescence (if any) because of rapid relaxation to the lower-energy emissive excitons[31-35] via carrier thermalization, exciton formation, and exciton thermalization[6]. These relaxation processes, combined with material properties, determine how the initial photoexcited state evolves and ultimately divides into the various emissive excitons near the optical bandgap.

At a given excitation energy in 1L-$WSe_2$, it is well established that the relative numbers of excitons, trions, biexcitons, and charged biexcitons depend on factors such as the density of initial photoexcitations, fermi level, encapsulation, lattice temperature, strain, and defect density[24-27]. Significantly less is known about the effects of different excitation energies, in part because the microscopic details of excited state thermalization in the presence of multiple excitonic species and relaxation pathways are complex[31, 34, 36]. Nevertheless, these processes are of deep fundamental interest[16, 34, 37, 38] and important for many technological applications[6].

Previous studies in 2D semiconductors[29, 33-35, 39] and 2D quantum well systems[40, 41] measured different formation rates of neutral excitons from an initial gas of excitons versus a higher-energy gas of unbound electrons and holes[33, 34]. These experimental investigations showed that the thermalization times are extremely fast (< 1 ps) in 1L-TMD semiconductors[32-34], which has been corroborated by multiple theoretical studies[31, 34]. However, these prior studies have only considered the influence of excitation energy on the formation rates of neutral excitons and have not addressed its effects on formation rates and yields of other species such as trions and biexcitons.

Here, we experimentally show that the generation of trions and charged biexcitons increases with increasing optical excitation energy in 1L-$WSe_2$. High-energy optical excitation is found to create greater populations of charged biexcitons and trions up to and through critical points of saturation. The energy dependence of the generation of these species is measured for energies that span the visible spectrum, revealing that trion and charged biexciton formation efficiency is maximized at energies greater than the quasiparticle bandgap. Combining the energy-dependent measurements with power- and temperature-dependent spectroscopies, two likely contributing mechanisms are identified. The first is enhanced trion and charged biexciton formation from a gas of unbound electrons and holes, and the second is thermalization to intermediate, long-lived states at the indirect bandgap. While future work is needed to deconvolve the roles of these mechanisms more quantitatively, our current observations lend important insights into the formation of higher-order exciton states. They identify phenomena that must be accounted for in microscopic models of thermalization, motivate consideration of multiexciton states in ultra-fast optical[33, 34, 42] and momentum-resolved[43] spectroscopies, and highlight how excitation energy influences the photogeneration of exciton complexes in 2D semiconductors.

## Results

*Enhanced trion and charged biexciton emission at elevated photoexcitation energy*

Figure 1 compares the photoluminescence spectra of 1L-$WSe_2$ under high- and low-energy photoexcitation (1.83 and 2.33 eV, respectively). The 1L-$WSe_2$ samples were mechanically exfoliated from a bulk crystal (HQ Graphene) followed by encapsulation in top and bottom layers of hBN (see methods), and unless otherwise specified, all measurements were conducted at cryogenic temperatures (<50 K). Figure 1a contrasts the photoluminescence spectra of the 1L-$WSe_2$ acquired at the different excitation energies while keeping the excitation density fixed (~$10^{11}$ $cm^{-2}$). In both spectra, we observe emission from the exciton (denoted 'X' in Fig. 1a), several charged exciton complexes, which include the inter- and intra-valley trions (T1 and T2)[44-46], and the negatively charged biexciton ($XX^-$)[25-27]. The assignments of these states are based on prior gate-dependent spectroscopy studies[26, 47-50] and confirmed by power-dependent measurements as in Figure 1b. We also note that the strong emission from the negatively charged trions and charged biexciton indicates that the sample is n-doped.

The relative intensities of the excitonic states in Figure 1a are significantly different between the two excitation regimes, despite being acquired at the same excitation density. At low excitation energy, emission from the trion and charged biexciton are *weaker* than the exciton, whereas, at high excitation energy, the integrated emission intensities of the trions and charged biexciton are both *greater* than the exciton. The higher-energy excitation generates ~1.3× more trions and ~5.8× more charged biexcitons than at the lower-energy excitation (at the specific temperature and excitation density in Figure 1a). These results demonstrate that photoexcitation of 1L-$WSe_2$ via the Rydberg resonances of the A and B excitons[4, 51-53] creates a different population of excitons than photoexcitation of high-energy exciton resonances (such as the C exciton) and/or the continuum of unbound free-carrier states[34, 52-54].

Because the charged biexciton state scales super-linearly with excitation density[26], its fraction of the emissive exciton population will increase with greater excitation density. To probe the combined effect of excitation density and energy, Figure 1c reports a separate measurement of the intensity of charged biexcitons relative to the neutral exciton for three different excitation energies (2.3, 2.0, and 1.9 eV) at excitation densities from ~$10^{10}$ $cm^{-2}$ ~$10^{12}$ $cm^{-2}$. For all energies, the relative intensity of the charged biexciton is maximized at an excitation density of ~$10^{11}$ $cm^{-2}$. At larger excitation densities, the (absolute) intensity of the charged biexciton saturates, whereas the exciton does not, causing the relative intensity of the charged biexciton to decrease. As shown in Figure 1c, the maximum fraction of charged biexciton emission before the point of saturation is 2.2× greater for excitation at 2.3 eV than at 1.9 eV, unambiguously demonstrating that high-energy excitation promotes increased formation of charged biexcitons relative to excitons in 1L-$WSe_2$. Figure 1d plots this maximum relative intensity of the charged biexciton for excitation energies from 1.87-2.34 eV, which exhibits a step-like transition from 2.0 – 2.1 eV. Finally, we note that the data in Figure 1a show a greater enhancement of the charged biexcitons than in Figures 1c and 1d. Similar differences in the enhancement factors are observed below. We attribute the different enhancement factors to variations in temperature between the two measurements (see below for temperature dependence) and/or age-induced changes in the doping/defect density of the 1L-$WSe_2$ (which is the target of future studies).

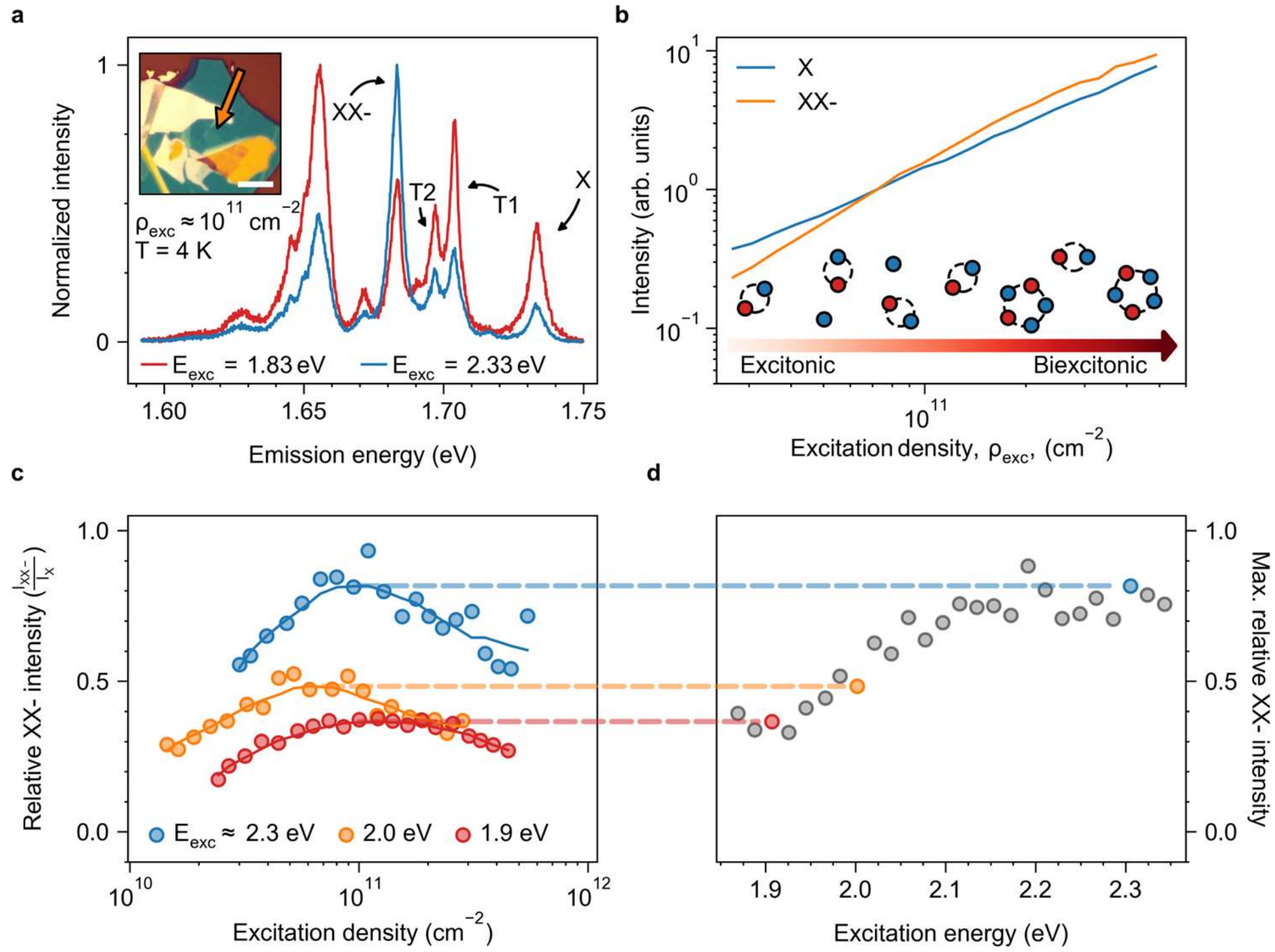

**Figure 1.** Enhanced relative emission of charged biexcitons with high-energy photoexcitation. (a) Two photoluminescence spectra of the hBN encapsulated 1L-$WSe_2$ sample (optical micrograph inset, scalebar: 10 μm) after optical excitation with low energy light, $E_{exc}$ = 1.83 eV (red), and high energy light, $E_{exc}$ = 2.33 eV (blue). (b) Excitation density-dependent photoluminescence spectroscopy of the exciton (X, blue) and charged biexciton (XX-, orange). (c) The power dependence of the charged biexciton emission relative to that of the exciton at three different excitation energies. The solid lines are moving averages of the data displayed by the scatter. (d) The saturation value of the relative intensity of the charged biexciton for 25 different excitation energies.

*Measurement of energy-enhanced generation of exciton complexes*

The results in Figure 1 show that the relative fraction of charged biexcitons increases at higher excitation energies, but only indirectly suggest that the emission from the charged biexciton is brighter for higher excitation energy at a fixed excitation density. Confirming that the charged biexciton is brighter requires careful control of the optical excitation process such that only the excitation energy is varied. To perform this absolute measurement, we implemented a modified photoluminescence excitation spectroscopy (PLE) technique that maintains constant excitation density and area for energies across the visible spectrum. With these parameters fixed, the photoluminescence intensity of each state is a measurement of the number of excitons, trions, and charged biexcitons that are generated in a manner that can be quantitatively compared for the different excitation energies. Alternatively, the excitation profile could be allowed to vary, and the photoluminescence intensities could be scaled to a constant excitation area for a given excitation density. However, the presence of states that depend nonlinearly on excitation density (e.g. super linear scaling or saturation behavior) and the vulnerability of different excitation areas to spatial disorder would introduce significant uncertainties[38] in such an alternative approach.

Figure 2 summarizes the key experimental parameters and the raw data of the modified PLE spectroscopy at fixed excitation density and area for 1L-$WSe_2$. In these measurements, the laser excitation energy is swept from 1.80 to 2.68 eV. The excitation profile is held constant for all excitation energies with a motorized and computer-controlled Gaussian telescope that tunes the collimation of the laser beam before it enters the optical microscope. As shown in Figure 2a, the average area of the photoexcitation spot size is 2.01 $\mu m^2$, corresponding to a FWHM of 1.60 μm, which is on average ~3× larger than the diffraction limit across the range of energies. To calibrate the density of excitations that are initially generated at a given energy and flux, both the absorptivity of the 1L-$WSe_2$ and the overall structure of the sample must be considered. The multilayer geometry of the sample (see the cartoon inset in Fig. 2a)[52] [55] modulates the photoexcitation external quantum efficiency (EQE), which is the fraction of incident light absorbed by the 1L-$WSe_2$, due to optical reflections at each interface in the hBN/1L-$WSe_2$/hBN/$SiO_2$/Si stack. To account for this effect and estimate the EQE at each photoexcitation energy, we use established transfer matrix method (TMM) calculations[56, 57] with plane-wave excitation and the optical response functions that are reported in prior studies for 1L-$WSe_2$ at 4 K[58], Si[59], $SiO_2$[60], and hBN[61]. Figure 2b shows that the TMM calculations capture the strong absorption resonances associated with the excitons in 1L-$WSe_2$ and a broadband envelope with a free spectral range of approximately 400 nm from the interference effects of the photonic stack. Discussions of this photonic model and the suitability of a plane wave approximation are provided in SI note 1.

Figure 2c presents the full suite of raw data of the modified PLE measurement. The photoexcitation energy steps in 8.8 meV increments, and the average excitation density ($\rho_{exc}$) at each energy is ~$3\times10^{11}$ $cm^{-2}$, equating to generating ~5000 excitations per laser pulse. Figure 2d shows three example PL spectra at the excitation energies of 1.83, 2.25, and 2.60 eV. The spectra are plotted without any additional normalization or scaling so that the absolute intensities of the excitonic species can be directly compared. At 1.83 eV photoexcitation, the exciton dominates, whereas at 2.25 eV and 2.60 eV, the trions and charged biexciton dominate, confirming the observations in Figure 1. To resolve how the emission intensity depends on excitation energy for a particular state across all energies, we integrate each PL spectrum around a small band centered on the respective emission feature. Figure 2e shows these integrated intensities at each excitation energy for the exciton (top panel), the sum of the inter- and intra-valley trions (middle panel), and the charged biexciton (bottom panel). A strong peak at 1.875 eV (discussed in detail below) on top of a constant background appears in the excitation spectrum of the exciton. For the trions and charged biexciton, a similar peak feature at 1.875 eV appears, but in strong contrast to the exciton, the peak is superimposed on a step-like increase in the emission intensity at ~1.95 eV, similar to that in Figure 1d.

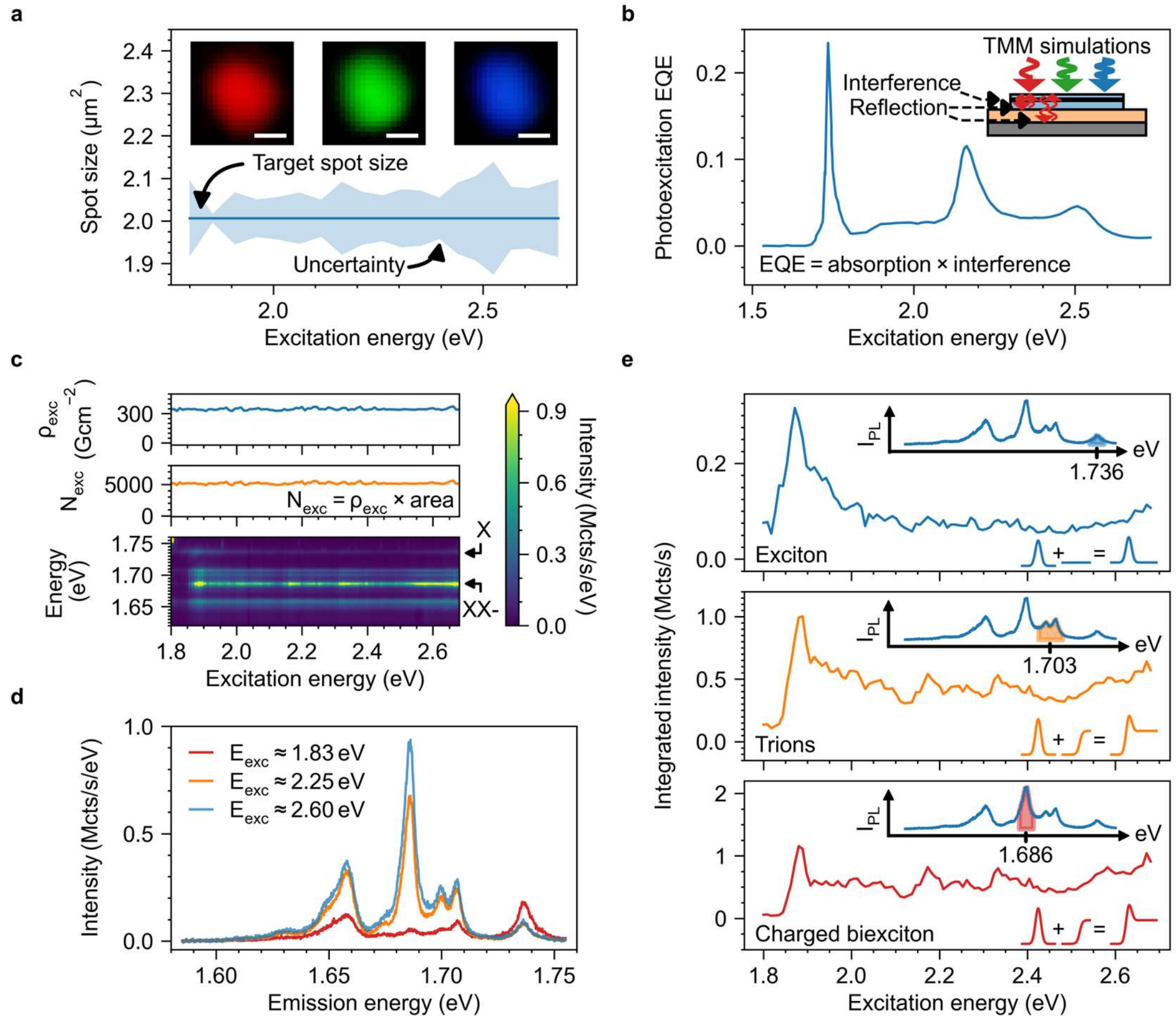

**Figure 2.** Microscopic excitation spectroscopy of linear and nonlinear states and enhanced trion and charged biexciton photoluminescence at elevated excitation energies. (a) Experimental control of the illumination area (i.e., spot size) for each excitation energy. The solid line indicates the target spot size, and the shaded region shows the estimated uncertainty. Insets are optical images of the focused laser spot on the sample using red, green, and blue laser colors. All scale bars are one μm. (b) The energy-dependent photoexcitation external quantum efficiency (EQE). The inset is a schematic of our sample's cross-section showing the 1L-$WSe_2$ encapsulated in hBN on a Si/$SiO_2$ wafer. The arrows illustrate the photonic cavity that modulates the EQE. (c) The photoluminescence spectra taken at 100 excitation energies ($E_{exc}$) from 1.80-2.66 eV while the excitation density ($\rho_{exc}$, top panel) and total number of excitations ($N_{exc}$, middle panel) are held constant for all energies. (d) Photoluminescence spectra at representative excitation energies from (c). (e) The excitation spectrum of the exciton (top panel), inter- and intra-valley trions (middle panel), and charged biexciton (bottom panel), extracted from the data in the bottom panel of (c). Upper insets: the colored overlays show the integration bands used to calculate the excitation spectra—bottom insets: schematics that separate the broadband trends from the potential fine structure feature that is dominated by the peak at 1.875 eV.

For all three species, the excitation spectra in Figure 2e exhibit an apparent fine structure. Most notably, the generation of all three states appears to be strongly enhanced in a narrow window of excitation energies around 1.875 eV. Figure 3 probes the nature and origins of these apparent resonances in more detail by considering the excitonic emission at low excitation densities where the neutral exciton dominates. In Figure 3a, the absorption spectrum of 1L-$WSe_2$ at 4 K reported in literature[58] (and used in our TMM calculations) is compared to the reflectance contrast and standard linear photoluminescence excitation (PLE) of the hBN/1L-$WSe_2$/hBN/$SiO_2$/Si stack studied here at 4 K. We note that the standard linear PLE spectrum reports the PL brightness as a function of excitation energy under a constant flux of photons (as opposed to a constant density of photoexcitations as probed in Figure 2). In the absence of changes in the neutral exciton generation rate, the linear PLE spectrum will follow the absorption spectrum at low excitation densities [52, 53, 62]. Indeed, the exciton PLE spectrum of our sample follows the absorption spectrum and is devoid of any features that indicate a dramatic change in the exciton generation rate, suggesting that the apparent fine structure in Figure 2e has different origins than enhanced generation rates.

Close comparison of the spectra in Figure 3a reveals that the $A_{2S}$ and $B_{1S}$ exciton resonances are more narrow in the PLE spectrum than in the reported absorption spectrum (a complete discussion of the features observed in the spectra is provided in SI note 2)[28, 63, 64]. From this comparison, we suspect that the reported absorptivity of 1L-$WSe_2$[58] underestimates the peak absorption of the excitonic resonances for our specific sample. To explore this hypothesis, we calculated a corrected 1L-$WSe_2$ absorptivity that keeps the integrated photoluminescence intensity constant for low excitation densities in the modified PLE measurement of Figure 2 (see SI note 3 for additional details). As shown in Figure 3b, this corrected absorption spectrum exhibits sharper exciton resonances -- especially for the $A_{2S}$ state -- that are more consistent with the linear PLE spectrum (Figure 3a). Such differences are not surprising since the hBN encapsulation is known to sharpen the exciton resonances [65].

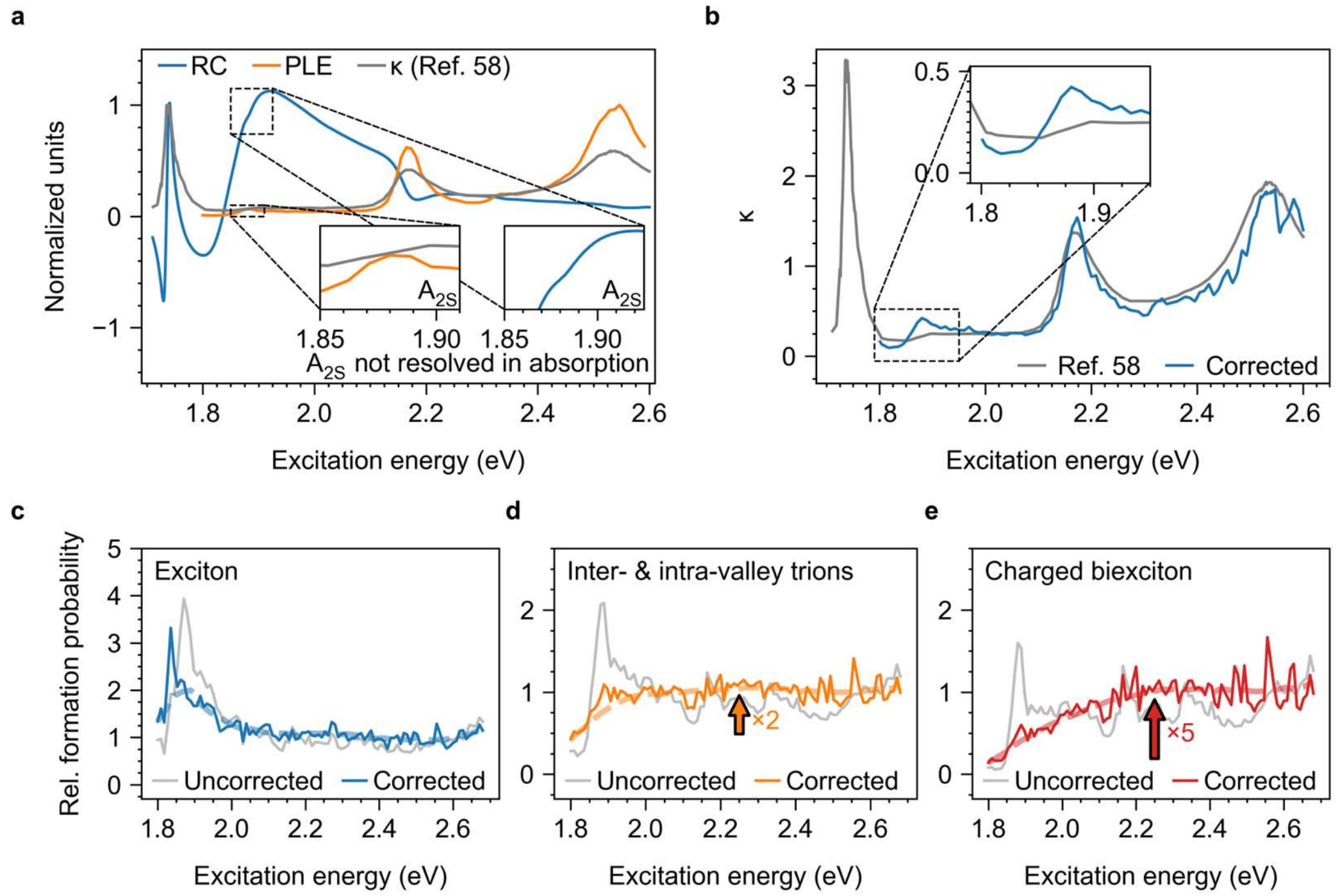

**Figure 3.** Analysis of the fine structure of excitation resonances and their elimination with a corrected absorption spectrum. (a) The measured reflectance contrast (RC, blue) and standard linear photoluminescence excitation spectroscopy (PLE, orange) compared to the absorption spectrum reported in ref. [58] (grey). (b) The extinction coefficient after applying a correction that maintains a constant photoluminescence quantum yield at low excitation densities (blue). The grey line is the uncorrected extinction coefficient. (c-e) The formation probabilities, relative to the saturation level at excitation energies greater than 2.5 eV, for the exciton, inter- and intra-valley trions, and charged biexciton. The corrected spectra are derived by enforcing that the photoluminescence quantum yield at low densities does not depend on the excitation energy. The colored lines show the corrected spectra, and the dashed lines show a moving average. Underlaid in grey is the uncorrected data.

Figures 3c-e show the resulting excitation spectra that were corrected to the differences in absorptivity. These spectra are our best estimates of how the formation probabilities of each state depend on excitation energy. Comparison to the raw data shows that the correction removes the fine structure features for the trion and charged biexciton, better resolves the step-like feature that marks an onset of increased formation probability for the trions and charged biexciton, and uncovers a slight reduction for the exciton generation as the excitation energy increases. Notably, the step-like increase in the formation probability of the trions and charged biexcitons now resembles the step-like feature in the maximum relative intensity of the charged biexciton shown in Figure 1d. All together, these trends provide unambiguous evidence that the excited state population shifts from excitons to a population of trions and charged biexcitons with increasing excitation energy. They also reveal that the efficiencies of trion and charged biexciton formation saturate to maxima at approximately 2× and 5×, respectively, at energies beyond 2.1 eV.

*Scaling of enhancement with excitation density and the potential role of hot free carriers*

From the corrected results of the modified PLE, the formation of charged biexcitons and trions at a given excitation density (~$10^{11}$ $cm^{-2}$) is most efficient at energies that are >400 meV above the optical bandgap. To extend and confirm the power-dependent characterization of Figures 1c and 1d, we investigated the power dependence of the enhancement at excitation densities spanning from ~$10^{10}$ $cm^{-2}$ to ~$10^{12}$ $cm^{-2}$ for the full range of excitation energies. Figures 4a-c show the (corrected) excitation spectra at three fixed densities. The enhancement effect is observed for all excitation densities with the same energy dependence as in Figure 3d. However, the magnitude of the enhancement of the trions and charged biexcitons at higher excitation energies depends on the density. For both states, the enhancement is maximal at an excitation density of about $10^{11}\ cm^{-2}$ as expected from the saturation discussed in Figure 1.

From the power-dependent measurements, the scaling relationship between the photoluminescence intensity ($I$) and excitation density ($\rho_{exc}$) is $I = I_0 \times (\rho_{exc})^{\alpha}$ can also be determined for the excitonic states at each excitation energy. Here, $I_0$ is a constant of proportionality, and $\alpha$ is a constant that characterizes the scaling relationship (i.e., $\alpha = 1$ for linear scaling, $\alpha > 1$ for superliner scaling, etc.). Figure 4d shows how the exciton photoluminescence intensity scales with excitation density for several excitation energies. The data in Figure 4d are shifted horizontally for clarity to show how the scaling relationship for the exciton evolves from linear to superlinear as excitation energy increases. Figure 4e shows the extracted power-law, $\alpha_X$, of the scaling relationship at each excitation energy. At the lowest excitation energies, the power-law is approximately linear ($\alpha_X \approx 1.02$), whereas at higher excitation energies, the power law is super-linear ($\alpha_X \approx 1.18$). Like the enhancement effect, the characteristic scaling of the exciton shows a step-like increase from low- to high-energy excitation, saturating at energies that are >400 meV above the optical bandgap.

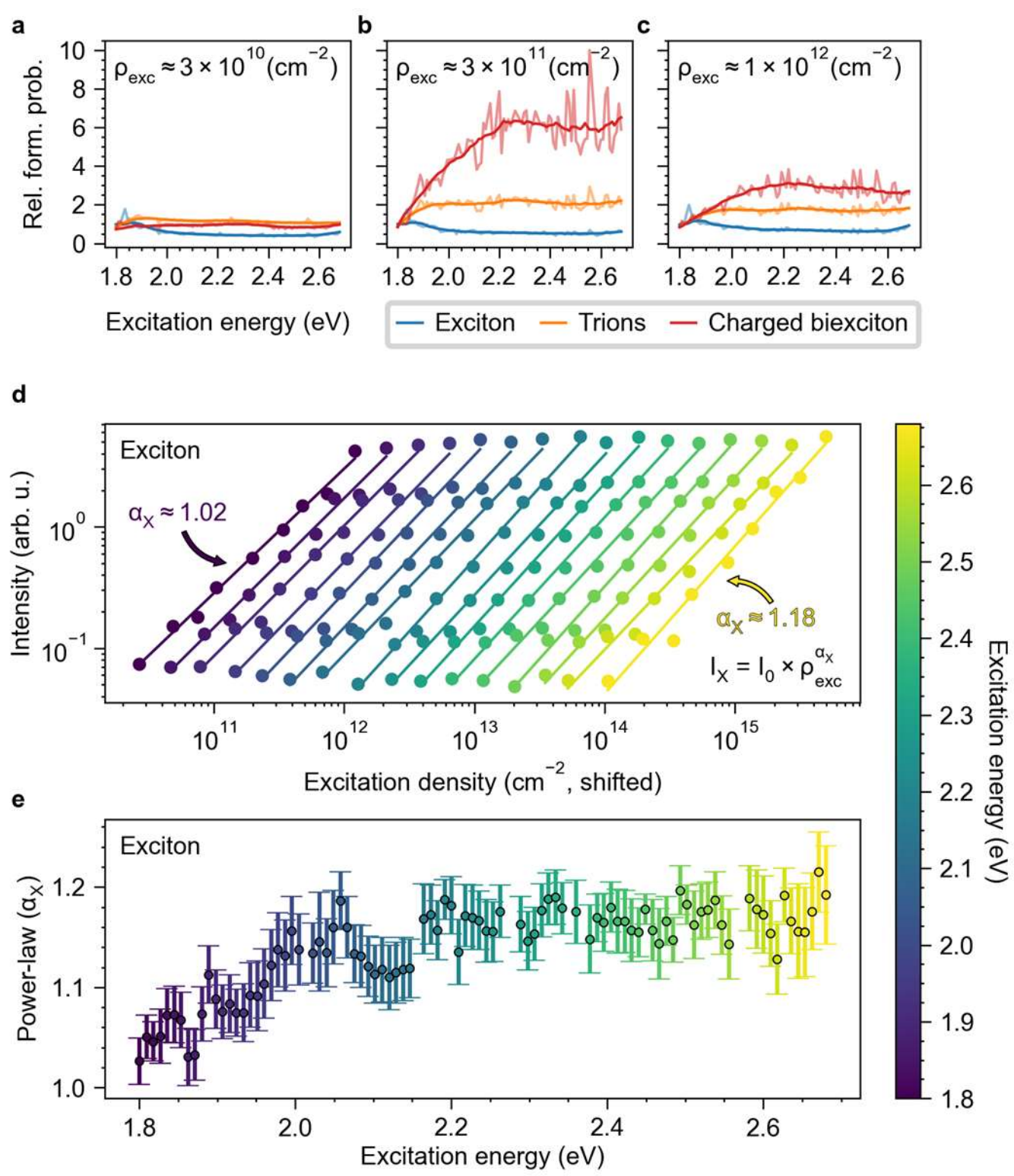

**Figure 4.** The density-dependence of the trion and charged biexciton enhancement and the power-law scaling relationship between the excitation density and exciton photoluminescence intensity. (a-c) The corrected excitation spectrum for the exciton (blue), trions (orange), and the charged biexciton (red) at increasing excitation densities scaled by their values at low excitation energy. (d) Examples of power-dependent photoluminescence intensity of the exciton with increasing excitation energy. Each measurement spanned the same range of excitation densities ($10^{10}$ $cm^{-2}$ - $10^{12}$ $cm^{-2}$) but is shifted horizontally for clarity. Each trend is fit to a power-law scaling relationship. (e) The excitation spectrum of the power law scaling of the exciton ($\alpha_X$) and estimated uncertainty.

The step-like transition of the exciton scaling law is in good agreement with the energy of the quasiparticle band gap (2.0 eV, see SI note 4 for details) and suggests that optical excitation of 1L-$WSe_2$ generally falls into one of two regimes: either above or below the quasiparticle bandgap. Below the quasiparticle bandgap, the direct photoexcitation of Rydberg excitons at the K-point efficiently populates the $A_{1S}$ exciton manifold due to the fast intra-valley relaxation rates in 2D semiconductors[33]. In this case, exciton formation is geminate, and the intensity of the exciton photoluminescence should scale linearly with excitation density. In contrast, higher energy photoexcitation initially populates the continuum beyond the quasiparticle bandgap[33-35, 39] and is more likely to form free carriers[34, 52, 54]. In this higher energy regime, the formation of excitons is a bimolecular process that scales as the product of the free electron and hole densities[66], analogous to the bulk recombination in other semiconductors[67, 68]. Depending on the formation efficiency, the power-law scaling in this high energy regime will be super-linear (see SI note 6 for a basic model). Therefore, the increase in the power-law reported in Figure 4e marks the transition between these formation regimes in the 1L-$WSe_2$, which is corroborated by dual-color pump-probe reflectivity measurements on our samples (see SI note 7 for details) and previous ultrafast spectroscopy[33, 34]. For completeness, SI note 8 reports excitation energy-dependent scaling of the trions, charged biexciton, defect band, and the total photoluminescence intensity. They all follow a similar trend across the quasiparticle bandgap, except for the defect photoluminescence. The similarity of the transition from below to above-gap excitation revealed by the exciton scaling law to the enhancement of the charged biexcitons and trions (Figs. 1d, 3c), qualitatively links the two phenomena: thermalization of above-gap excitations may contribute to the increased generation of charged biexcitons and trions at high excitation energies.

*Temperature dependence of the enhancement effect*

Finally, to probe for a critical energy scale of the processes that underlie the enhancement effect, Figure 5 characterizes how the energy-induced enhanced generation of trions and charged biexciton states depends on temperature up to 80 K. Above 80 K, the photoluminescence from the trions and biexcitons is suppressed, which has been reported in prior work[47-50, 69] and prohibits measuring the enhancement at higher temperatures. In Figures 5a and 5b, the temperature-dependent photoluminescence of the 1L-$WSe_2$ is reported for low-energy excitation (1.88 eV) and high-energy excitation (2.36 eV), both at an excitation density of $10^{11}$ cm$^{-2}$. In this dataset, the FWHM sizes of the illumination areas are 1.04 μm for 1.88 eV and 0.84 μm for 2.36 eV. To account for the resulting differences in total excitations, the PL spectra in Figures 5a and 5b are scaled such that the exciton emission at 4 K has a peak intensity of unity (which is equivalent to scaling by the total areas as discussed above). Comparing the scaled intensities of the charged biexcitons at the two excitation energies, it is evident that their enhancements persist to 80 K but with significantly diminished magnitudes than at 4 K.

To capture the suppression of the effect at elevated temperature, the enhancement for the trions and charged biexcitons are calculated at each temperature from the ratio of the scaled intensities under photoexcitation at 2.36 eV to that under photoexcitation at 1.88 eV. In terms of the data in Figures 5a and 5b, the enhancement for the charged biexciton ($\Gamma_{XX-}$) at each temperature is determined by,

$$\Gamma_{XX-} = \frac{I_{2.36\,eV}^{XX-} \Big/ I_{2.36\,eV}^{X;\,T\,=\,4\,K}}{I_{1.88\,eV}^{XX-} \Big/ I_{1.88\,eV}^{X;T\,=\,4\,K}}$$

The enhancement of the trions ($\Gamma_{trions}$) is calculated in a similar manner, and the temperature dependence of both quantities is shown in Figure 5c. The enhancement factors for both states decrease as the temperature exceeds 10 K and converges to unity at higher temperatures. To find a characteristic activation energy, we fit the temperature dependence of the enhancements to an Arrhenius model,

$$\Gamma = \frac{\Gamma_0}{1 + A \times exp\left(^{-E_a}\big/_{k_b T}\right)}$$

where $\Gamma_0$ is the enhancement factor at 0 K, $E_a$ is the characteristic activation energy of the process (or processes) that suppress the enhancement, $A$ is a free parameter corresponding to the low-temperature generation rates, and $k_b$ is the Boltzmann constant. This model quantitatively captures the temperature dependence of the enhancement effect and yields an activation energy of around 8-9 meV (~100 K) for both the trions and charged biexciton, which is 2-3× smaller than the binding energies of the states[48-50, 69].

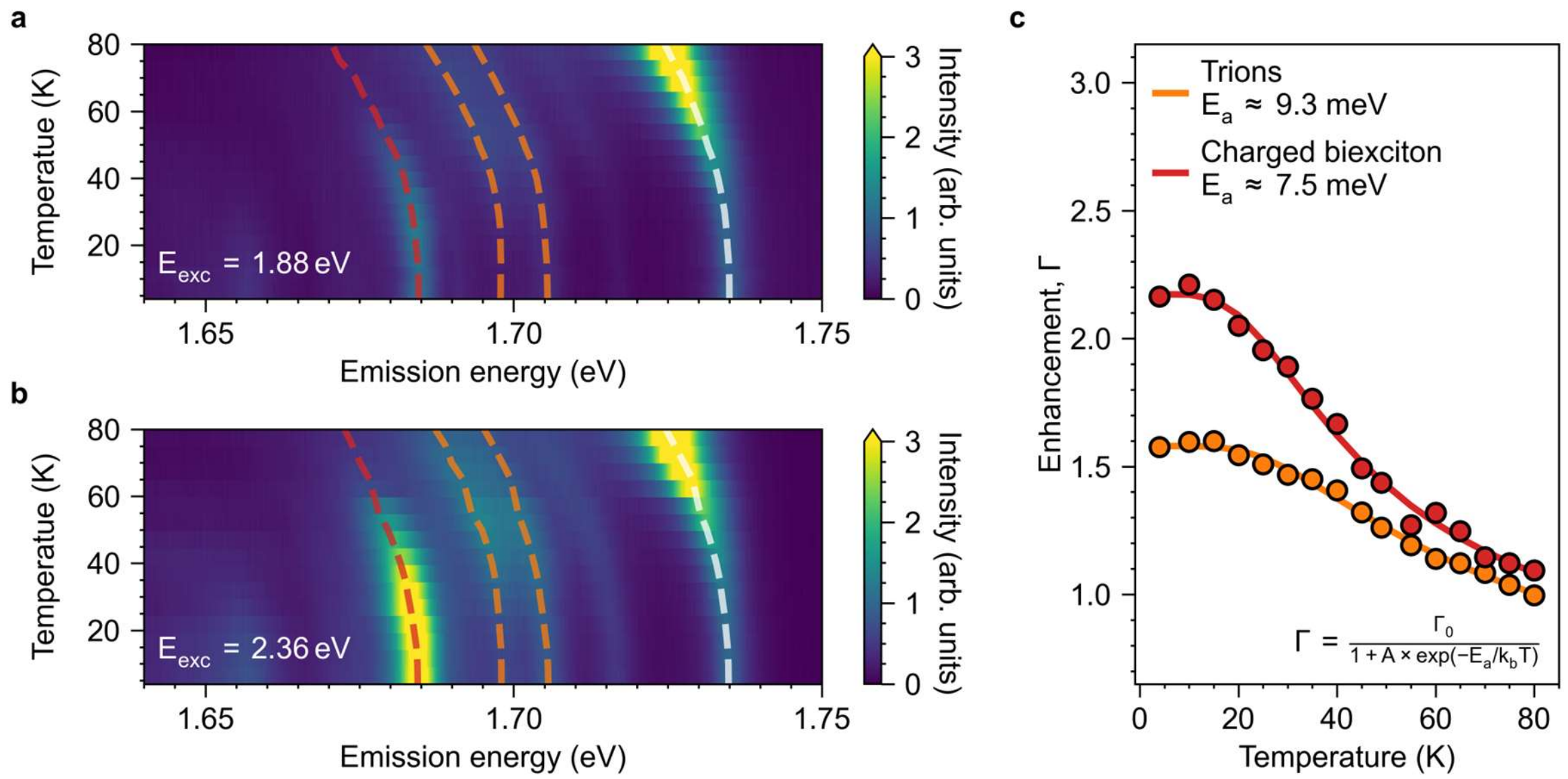

**Figure 5.** Temperature dependence of the enhanced formation of trions and charged biexcitons. (a), (b) Temperature-dependent photoluminescence at 1.88 eV and 2.36 eV, respectively. The excitation density is held constant at about $10^{11}$ $cm^{-2}$. The dashed lines track the emission energy of the exciton (white), trions (orange), and charged biexciton (red) as the temperature of the sample changes. (c) The temperature dependence of the enhancement factor for the trions (orange, scatter) and charged biexciton (red, scatter). The data is fit to an Arrhenius equation to find the characteristic activation energy of the process (or processes) that suppress the enhancement effect, as shown by the colored lines.

## Discussion

The results shown above demonstrate that excitation above the quasiparticle band gap in 1L-$WSe_2$ can enhance the formation of trions and charged biexcitons. To the best of our knowledge, this effect has not been previously reported in any system nor treated in any theoretical model. From the data here, we can restrict the potential underlying mechanism of the enhancement from a broader set of possibilities. The excited state dynamics reported in SI note 9 show that the increased photoluminescence intensity of the charged biexciton does not correspond to a change in its excited state lifetime, confirming that the enhancement manifests in the formation dynamics. The temperature dependence shows that the thermal energy needed to suppress the enhancement is significantly smaller than the binding energies of the exciton complexes (15 meV – 30 meV)[24-27, 70] and the dark-bright exciton fine structure splitting at the K-point (30 meV)[5, 17, 18, 20, 44], excluding significant contributions from the dissociation of excitons or the temperature-induced brightening of K-point dark excitons. Further, this thermal energy equates to a temperature of about 100 K, vastly exceeding the critical temperatures of correlated ground states, such as Wigner crystallization in 1L-$MoSe_2$[71]. In contrast, the activation energy is closer to the predicted energetic splitting between the momentum-indirect KQ exciton and spin-bright KK exciton (~6 meV)[72] and the binding energies of localized excitons in TMD semiconductors (10 – 100s meV)[73, 74].

Photo-induced changes in the carrier density of the 1L-$WSe_2$ could also play an important role. Previous experiments have established that the formation of trions and charged biexcitons depends on the static background charge density[24-26]. Furthermore, there is evidence that photodoping can arise from photoionizing color centers in the supporting hBN crystallites[75] and interfacial charge transfer. To control for the possibility that our sample accumulates a static charge during the measurements, we modulated the photoexcitation between 1.82 and 2.36 eV. These controls show that the absolute intensities of the states are highly repeatable over multiple cycles of the excitation energy (see SI note 10), excluding any quasi-persistent charging of the sample.

Consequently, based on prior studies, our current data, and our analysis, we hypothesize that two mechanisms contribute to the enhancement effect: (1) enhanced formation of KQ excitons at high-energy photoexcitation, which has been observed in recent time- and angle-resolved photoemission spectroscopy measurements and (2) the direct formation of excitonic states from the photogenerated free carrier gas. Detailed consideration of master rate equations and more advanced microscopic thermalization models that include these two potential mechanisms motivates future studies at higher excitation densities (ideally up to the excitonic Mott transition) and probing/correlating biexciton states with momentum-resolved spectroscopies such as time-resolved ARPES[43]

Additionally, we note that in light of the spectroscopy presented here, the extent to which the formation of KQ excitons at large photoexcitation energies[43] alters the total quantum yield of 1L-$WSe_2$ remains an open question. Our spectroscopy suggests that the quantum yield subtly increases at elevated excitation energies and densities, perhaps due to the brightening of the excited-state population from the enhanced formation of charged biexcitons. However, to unequivocally resolve this question, the accuracy of the photoexcitation EQE calibration needs to be improved, especially near the $A_{2S}$ absorption resonance. The doping level should be controlled for these measurements, and the spin-dark exciton population at the K-point should be measured, perhaps using an edge-on collection geometry[76]. An accurate dielectric function for the 1L-$WSe_2$ under these conditions is essential here (as well as in other quantitative experiments), necessitating improvement of Kramers-Kronig analysis of reflectance contrast spectra for encapsulated samples on non-transparent substrates.

In conclusion, we have shown that the photoexcitation energy affects the formation of exciton complexes in 1L-$WSe_2$. Energies above the quasiparticle bandgap substantially enhance the trion and charged biexciton populations. To the best of our knowledge, such observations have not been previously reported and are presently unique to 1L-$WSe_2$. The enhancement is uncovered with excitation spectroscopy where electromagnetic modeling and a uniform illumination area are used to keep both the excitation density and the total number of excitations constant over a broadband range of excitation energies. This technique offers an alternative to photoluminescence quantum yield measurements[77] to probe the energy-dependent formation probabilities of linear excitations (excitons and trions) and nonlinear excitations (charged biexcitons) in microscopic samples. Combined with both temperature- and power-dependent spectroscopies, we identify that mechanisms that may underlie the effect (and warrant additional theoretical and experimental investigation) include the formation of momentum-dark excitons, carrier capture by defects, and favored formation of charged exciton complexes from an electron-hole gas. While the unambiguous determination of the underlying mechanism will require future study, the enhancement highlights how the complexity of excited state thermalization and exciton formation in 2D materials at low temperatures can tailor the resulting sub-populations of excitons responsible for light emission and other optoelectronic processes.

## Methods

### *Sample fabrication*

We assembled the fully encapsulated hBN/1L-$WSe_2$/hBN sample by first exfoliating 1L-WSe2 and thin hBN crystallites using scotch tape, then aligning the constituent crystallites onto a polymer stamp, and finally transferring them to the target Si/$SiO_2$ substrate using a flip-chip technique[78]: each layer of the heterostructure stack was sequentially picked up in bottom-up order by a polymer stamp (PPC + PDMS) before flipping the stack onto the supporting $Si/SiO_2$ substrate. We vacuum-annealed the sample to remove the underlying PPC polymer after assembly.

*Cryogenic photoluminescence spectroscopy*

We conducted the cryogenic photoluminescence measurements by cooling the sample in a closed-cycle liquid helium cryostat with optical access (Montana Instrument s50 Cryostation) and illuminating the sample with pulsed laser light generated from a supercontinuum fiber laser (NKT photonics SuperK EXTREME, pulse width = 55 ps, rep. rate = 78 MHz) filtered with an acoustic-optic tunable filter (Gooch and Housego, filter bandwidth ≈ 4 nm). Light was focused onto the samples using a Nikon 40× objective (NA = 0.6) with tunable correction for spherical aberration. Luminescence was collected in a back-scattered geometry and directed to a Czerny-Turner optical spectrometer (Horiba iHR320), dispersed with a mechanically ruled grating (groove density = 600 g/mm), and measured with a scientific CCD camera (Andor iDus 416, cooled to -55 °C) after rejecting laser light with a series of interference filters. We used a 4f-imaging system and a 2-axis galvanometric scan mirror to position the laser on the sample.

*Calibration of the illumination area for quantitative photoluminescence excitation spectroscopy*

We used a Gaussian telescope to tune the collimation of the laser to maintain a constant illumination area across all excitation energies. We monitored the illumination profile while we changed the separation of the lenses and the laser wavelength to find the optimal configuration of the telescope: we took images of the focused laser spot as we changed the color of the laser and varied the lens separation. Each image was fit with a 2D Gaussian intensity profile. For our experiments, we selected the smallest spot area our system could maintain across the extent of the excitation energies probed (1.80 eV – 2.68 eV), which had an area of about two $\mu m^2$. We estimate that the uncertainty in the spot area is +/- 3.3% due to the coarseness between points in the laser color/lens separation grid used to generate the calibration curve.

*Definition of the excited state density*

In our experiments, we define the excited state density as

$$\rho_{exc} = \frac{(P[J/s]) \times (6.242 \times 10^{18}[eV/J]) \times (EQE)}{(78 \times 10^{6}[pulses/s]) \times (E_{exc}[eV]) \times (A[cm^{2}])}$$

where $P$ is the time-averaged power delivered to the sample by the incident laser, $6.242 \times 10^{18}[eV/J]$ is the conversion from Joules to eV, EQE is the photoexcitation external quantum efficiency, $78 \times 10^{6}[pulses/s]$ is the repetition rate of the laser, $E_{exc}$ is the excitation energy in eV, and A is the characteristic illumination area. We note that this definition reports the excitation density integrated throughout each laser pulse (55 ps, FWHM) and that the instantaneous excitation density will depend on the dynamics of the photoexcitations.

*Linear photoluminescence excitation spectroscopy*

We measured linear PLE by maintaining a constant photon flux through the monolayer while accounting for thin-film interference effects using the TMM model.

*Reflectance contrast spectroscopy*

The reflectance contrast was measured by imaging a back-illuminated pinhole onto the sample and calculating the normalized difference between the reflectance spectrum in regions consisting of hBN/1L-$WSe_2$/hBN/$SiO_2$/Si versus hBN/$SiO_2$/Si. The reflectance contrast here is,

$$\mathrm{RC} = \frac{R_{\mathrm{off}} - R_{\mathrm{on}}}{R_{on} + R_{off}}$$

Where $R_{on}$ is the reflection from the regions that include 1L-$WSe_2$.

ASSOCIATED CONTENT

**Supporting Information**

Saturation behavior of the charged biexciton at elevated excitation densities, transfer matrix method simulations used to calibrate the photoexcitation external quantum efficiency, discussion of the low-energy resonance in the excitation spectra, the procedure to estimate the uncertainty of the power-law fitting, estimation of the quasiparticle bandgap energy, basic model of power-law suppression, additional evidence for the optical generation of free carriers by dual-color transient reflection spectroscopy, trion and charged biexciton power laws, excitation energy-dependent relaxation dynamics of the charged biexciton, and repeatability of the enhancement effect.

AUTHOR INFORMATION

**Corresponding Authors**

Nicholas J. Borys - Department of Physics, Montana State University, Bozeman MT, 59718. Email: nicholas.borys@montana.edu.

**Present Addresses**

[†]Matthew C. Strasbourg - Department of Mechanical Engineering, Columbia University, New York, NY

**Author contributions**

N.J.B., M.C.S., and P.J.S. conceived this work. C.J. and Z.N. built instrumentation to automate the power-dependent spectroscopies. E.Y. and T.P.D. fabricated the hBN/$WSe_2$/hBN heterostructure stack and provided initial characterization data. M.C.S. conducted the optical spectroscopy measurements, data analysis, and subsequent additional characterization of the sample. M.C.S., P.S., S.A., and E.M.G. fabricated samples and performed ultrafast characterization measurements. All authors contributed to the writing of the manuscript.

**Notes**

Data availability -The authors declare that all data supporting the findings of this work are available from the corresponding authors upon reasonable request.

Code availability - The authors declare that all analysis code supporting the findings of this work are available from the corresponding authors upon reasonable request.

Competing interests - The authors declare no competing interests.

ACKNOWLEDGMENTS

We thank C. Crites and J. P. Fix for their assistance with the atomic force microscopy measurements and A. Vorontsov for helpful discussions. N.J.B. and P.J.S. acknowledge support from the National Science Foundation through awards NSF-1838403 and NSF-2004437. This work was performed in part at the Montana Nanotechnology Facility, a member of the National Nanotechnology Coordinated Infrastructure (NNCI), which is supported by the National Science Foundation (Grant# ECCS-2025391). N.J.B. and P.S. acknowledge support from the MonArk NSF Quantum Foundry supported by the National Science Foundation Q-AMASE-i program under NSF award No. DMR-1906383. Development of select encapsulated devices used as part of these studies was supported by Programmable Quantum Materials, an Energy Frontier Research Center funded by the US DOE, Office of Science, Basic Energy Sciences (BES), under award DE-SC0019443.

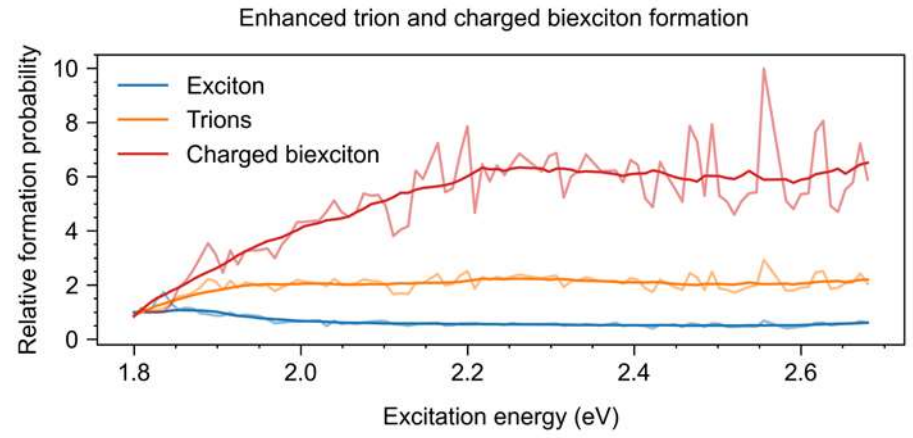

For table of contents only.